\documentclass[letter, twocolumn,superscriptaddress, floatfix, nobalancelastpage]{revtex4}
\usepackage{amsfonts}
\usepackage{amssymb}
\usepackage[dvips]{graphicx}
\usepackage{color}
\usepackage{amsmath}
\usepackage[bookmarksnumbered, bookmarks, breaklinks, linktocpage]{hyperref}

\begin{document}

\title{N\'eel order in the two-dimensional $S=\frac{1}{2}$-Heisenberg Model}
\author{Ute L\"ow}
\affiliation{Theoretische Physik, Universit\"at zu K\"oln, Z\"ulpicher
 Str.77, 50937 K\"oln, Germany}

\date{\today}

\begin{abstract}
The existence of N\'eel order in the $S=\frac{1}{2}$ 
Heisenberg model on the square lattice  at $T=0$ is shown
using inequalities set up by Kennedy, Lieb and Shastry in combination with high precision 
Quantum Monte Carlo data. 

\end{abstract}

\maketitle

The ground state order of quantum spin systems, 
in particular the issue whether the ground state shows long range magnetic order, 
has attracted long and continuous interest.
For the prototype of spin models, the antiferromagnetic Heisenberg model,
the existence of N\'eel order  at low temperatures 
was proved 
in the seminal paper of  Dyson, Lieb and Simon \cite{DLS} in 1978
for spin $S\geq 1$ and spatial dimension
$d\geq3$  and also for $S=\frac{1}{2}$ and $d>3$.

Ten years later  Kennedy, Lieb and Shastry \cite{KLS} showed that
also for $S=\frac{1}{2}$ and $d=3$   N\'eel order in the
ground state exists.

The situation in two dimensions is different and more subtle, since the Mermin-Wagner-Hohenberg
theorem forbids N\'eel order at finite $T$, leaving open however the possibility of
N\'eel order in the ground state.
The existence of N\'eel order 
for the two-dimensional model and $S\geq 1$ 
was shown in \cite{AKLT,NP} and later in \cite{KLS} by an
independent derivation of the relevant inequality at $T=0$.

However the inequalities sufficient to show N\'eel order for $S=1$ 
in the two-dimensional case are not sufficient
to construct an analogous proof for $S=\frac{1}{2}$.
Thus the case of $S=\frac{1}{2}$ remains an open problem.
Still it is possible to derive inequalities 
concerning spin-spin correlations at {\sl short} distances \cite{KLS} which
are violated if N\'eel order is present.
That is, with a minimum of numerical information,
the question of N\'eel order in the ground state can be decided.

The issue of this paper is to evaluate the spin-spin correlations 
of the two-dimensional $S=\frac{1}{2}$ antiferromagnetic Heisenberg model
at {\sl short} distances
and demonstrate  that these results combined with the analytic expressions of \cite{KLS} 
show the existence of 
N\'eel order in the two-dimensional
$S=\frac{1}{2}$ antiferromagnetic Heisenberg model at $T=0$.
Such a study has become possible, due to the developement of high precision
Monte-Carlo techniques over the last decade.

In Ref.\cite{KLS}  Kennedy, Lieb and Shastry  used data of Gross,
Sanchez-Velasco and Siggia \cite{GSS}
for a comparison, however 
these data clearly deviate from the results presented here.
The authors of  \cite{GSS} used a Quantum Monte Carlo method without loop updates and  with discrete
Trotter time (see below). Their data served only as a crude comparison to 
extrapolated Lanczos data and data produced by the Neumann-Ulam method, which
were the best algorithms to study the
properties of the two-dimensional Heisenberg model in 1988.  
Today modern loop algortihms by far outreach both methods.

As will be shown in the following an accurate evaluation of correlation
functions at short distances is possible  with modern
Quantum Monte Carlo methods, which allow us to compute expectation values
at very low temperatures and even though the short distance results have a certain
finite size and finite temperature correction, these uncertainties are well controlled and 
allow  to draw definite conclusions.

The approach and intention of this paper is diffrent from a completely
numerical 
evaluation of e.g. the correlation length,
which involves a calculation of correlations at {\sl long} distances and an appropriate
extrapolation to {\sl infinite} distances, which cannot 
be used as a proof of long range order in any rigorous sense.

At first sight a "Quantum Monte Carlo algorithm" seems a puzzling concept, 
since an important step in any Monte-Carlo-method is the
evaluation of Boltzmann weights for given energies of the system.
For quantum models these energies are hard if not impossible to calculate. 
A key idea to make Monte Carlo methods applicable to
quantum systems is to map the quantum model onto a classical
model by introducing an extra dimension, usually referred to as
Trotter-time \cite{SUZ}.

In the first generation of algorithms this mapping was 
straightforwardly applied to the quantum Heisenberg model.
Though this allowed for a wealth of new studies of the finite
temperature properties in one and in particular in two-dimensional systems,
these algorithms had two major drawbacks, which became most evident at
low temperatures. Firstly the extra Trotter dimension was discretized, 
introducing the number of time slices as a parameter 
which had to be eliminated from the final results 
by an extrapolation. 
Secondly the update procedure, i.e. the construction of new
independent configurations, was done locally.
As a consequence one had to move through the lattice site by site
several times to obtain a configuration independent of the starting
configuration and useful for a new evaluation of an observable.

A first improvement was introduced by the so called
loop-algorithms \cite{ELM}, which uses nonlocal updates similar to the Swendsen-Wang 
algorithm for classical models. 
A second and important step towards high precision 
Quantum Monte Carlo techniques were algorithms which work directly in the 
Euclidian time continuum \cite {FG} and  require no extrapolation
in Trotter time. For the algorithm \cite{BW} used for the analysis presented here
no approximations enter, and statistical errors are the only source of inaccuracy.

Since this work intends to produce highly accurate data it seems
appropriate to assess the precision of the method by a comparison with exact
results. The best candidate for such a comparison are the correlations
of one-dimensional systems evaluated by the Bethe-ansatz
with almost arbitrary precision up to distance seven \cite{SST}
and with results for finite chains from Ref.\cite{DGHK}.
This is done in the Appendix for chains of 400 sites at T=0.005.

After these introductory remarks  we now  return to our actual goal, 
which is the two-dimensional system. 
Our starting point is a $S=\frac{1}{2}$ Heisenberg model 
\begin{eqnarray}
\label{eq:0}
H=\sum_{\underset {x,y \varepsilon \Lambda} {<\large xy>}} \vec S_x \vec S_y
\end{eqnarray}
with nearest neighbour interaction
on a finite square lattice $\Lambda$ with an
even number of sites in every direction and periodic boundary conditions.

The Fourier transform of the spin-spin correlation function at $T=0$ is given by
\begin{eqnarray}
\label{eq:1}
g_q=\langle S_{-q} S_{q} \rangle =\sum_{x\varepsilon \Lambda} e^{-iqx} \langle S_0^3 S_x^3 \rangle
 \end{eqnarray}
where 
\begin{eqnarray}
\label{eq:2}
S_q= \frac{1}{\sqrt {|\Lambda|}}  \sum_{x\varepsilon \Lambda} e^{-iq x } S^3_x.
\end{eqnarray}
For the corresponding finite temperature expectation value of $g_q$ an upper
bound $f_q$ was 
derived in \cite{DLS}. The $T=0$ limit of this bound 
was obtained in Ref.\cite{NP} and a direct proof of the bound at $T=0$ was given in \cite{KLS}.
Following the notation and arguments of 
Kennedy, Lieb and Shastry \cite{KLS} 
 the inequality for $d=2$ reads
\begin{eqnarray}
\label{eq:3}
g_q \leq f_q \ \ \ \ {\text for } \ \ \ q\neq Q
\end{eqnarray}
where 
$f_q=\sqrt\frac{e_0 E_q}{12 E_{q-Q}}$, $E_q=2-\cos q_1-\cos q_2$,
$Q=( \pi,\pi)$ and $-e_0$ is the ground state energy per site of the Heisenberg model
Eq.\ref{eq:0} on the lattice $\Lambda$.

The fundamental idea is, that the existence of N\'eel order in the 
limit of infinite system size corresponds to a 
delta-function in the Fourier transform of the spin-spin correlation
$g_q$ at Q. 
This means, if Eq. \ref{eq:3} is integrated over the whole Brillouin
zone one finds in the case of N\'eel order  
\begin{eqnarray}
\label{eq:4}
m^2+\int d^2q \ f_q \geq  \int d^2q \  g_q =S(S+1)/3 
\end{eqnarray}
where $m^2$ is the coefficient of the delta-function at $Q$.
%and 
%\begin{eqnarray}
%\int d^d q =\frac{1}{(2\pi)^d}\int_{0}^{2\pi} d q_1 \dots \int_{0}^{2\pi} d q_d .
%\end{eqnarray}

If there is no N\'eel order $m^2$ is zero. 
By numerically evaluating the integral over $f_q$,
and by using exact variational upper and lower bounds on 
the ground state energy $-e_0$
one sees, that the above inequality and its analogon for $d\ge3$
cannot be fulfilled with $m^2 =0$ and $S\geq1$, which proves 
N\'eel order.

Inequalities of type Eq. \ref{eq:4} are not sufficient to prove the
existence of a nonzero  $m^2$ for $d=2,3$ and $S=\frac{1}{2}$, but a new relation is obtained
by multiplying  $g_q$ 
by $\cos q_i$ and again integrating over
the Brillouin zone:
\begin{eqnarray}
\label{eq:5}
 \int d^d q  \  g_q \ \cos
q_i = \langle S_0^z S_{\delta_i}^z \rangle =-e0/3d 
\end{eqnarray}
with i=1,2 for d=2 and i=1,2,3 for d=3 and $\delta_i$ the unit vector in
i-direction and the value of the ground state energy form Ref.\cite{Sand} is $e_0=0.669437(5)$.

Carrying out an analogous integral over $f_q$ and using again Eq.\ref{eq:3} 
one finds:

\begin{eqnarray}
\label{eq:6}
\frac{e_0}{3d} \leq -\sqrt{ \frac{e_0}{6d}} \int d^dq
\sqrt{\frac{E_q}{d^2 E_{q-Q}}} \left(\sum_{i=1}^d \cos q_i\right)_+
\end{eqnarray}
were the $f_+$  means the positive part of a function, which equals f, when f is positive
and is zero otherwise.

Again Eq.\ref{eq:6}, which is valid if no N\'eel order  exists,
was shown to be violated for $d=3$ and $S=\frac{1}{2}$ in Ref.\cite{KLS} by using bounds on
$e_0$  and thus the existence of N\'eel order was proved also
for $d=3$ and $S=\frac{1}{2}$.

For $S=\frac{1}{2}$ and $d=2$ one cannot construct a contradiction
by using only the ground state energy. Here more input from numerical data is needed.
This can be incorporated by multiplying 
$g_q$ by $\cos(mq_i)$ 
with $m=2,3...$ 
and again integrating over the whole Brillouin zone:

\begin{eqnarray}
\label{eq:7}
\int d^2 q  \  g_q \ \cos(mq_i)  
= \langle S^3_0 S^3_{m \delta_i} \rangle    
\end{eqnarray}

with i=1,2. 

Next, defining $\bar g(n)$   as
\begin{eqnarray}
\label{eq:8}
\bar g(n) = \frac{1}{n+1} \sum_{m=0}^n (-1)^m \langle S_0^3 S^3_{m \delta_i} \rangle
\end{eqnarray}

and using again inequality \ref{eq:3}
one constructs the following  relations involving the correlation
functions: 
\begin{eqnarray}
\label{eq:9}
\bar g(n) = \int d^2 q \frac{1}{2n+2}\sum_{m=0}^{n} (-1)^m
\{\cos(mq_1)+\cos(mq_2)\}\ g_q \nonumber \\
\le \int d^2 q \frac{1}{2n+2}\sum_{m=0}^{n} (-1)^m
\{\cos(mq_1)+\cos(mq_2)\}_{+}\ f_q.\nonumber \\ 
\ 
\end{eqnarray}

Whenever the inequality  Eq. \ref{eq:9} is violated for a certain n, a nonzero $m^2$ 
multiplying the delta-function at $Q$ is needed and therefore the existence of 
N\'eel order is proved.

\begin{figure} 
\begin{center}
\includegraphics[width=8cm]{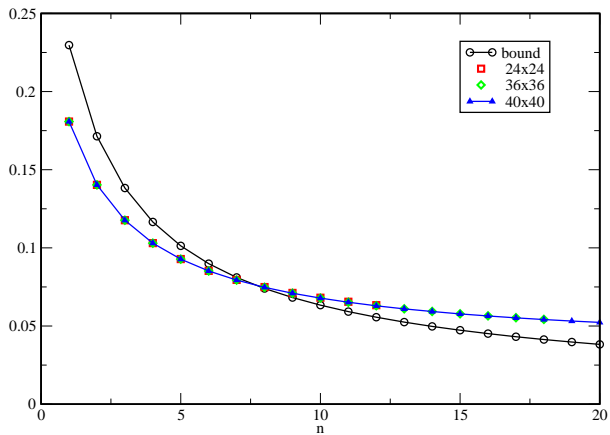}
\end{center}
\caption{Bound on $\bar g(n)$ obtained from Eq.\ref{eq:8} and $\bar g(n)$ for
$24\times 24$, $36\times36$ and $40\times40$ at $T=0.025$.}
\label{Fig}
\end{figure}

The $\bar g(n)$ as defined in Eq.\ref{eq:8} were calculated 
by the Quantum Monte Carlo method \cite{BW}.
The results, displayed in  table \ref{table1}, show that the $\bar g(n)$ calculated by 
Quantum Monte Carlo cross the bound obtained by integrating over $f_q$ at $n=8$.
This is also depicted in Fig. \ref{Fig}.
Thus inequality Eq.\ref{eq:9} is violated and  N\'eel order must exists in the two-dimensional
antiferromagnetic Heisenberg model with $S=\frac{1}{2}$ at $T=0$. 

\begin{table}
\begin{center}
\begin{tabular}{c|c|c|c|c|}
        n& Bound & $T=0.005$ & $T=0.025$ & $T=0.075 $  \\\hline
   1 &  2.297e-01    &  1.80799e-01  $\pm$  3.63e-06 & 1.80794e-01 &      1.80792e-01 \\
   2 &  1.714e-01    &  1.40308e-01  $\pm$  5.63e-06 & 1.40302e-01 &      1.40298e-01 \\
   3 &  1.383e-01    &  1.17686e-01  $\pm$  6.84e-06 & 1.17678e-01 &      1.17670e-01 \\
   4 &  1.166e-01    &  1.03005e-01  $\pm$  7.67e-06 & 1.02997e-01 &      1.02985e-01 \\
   5 &  1.013e-01    &  9.27815e-02  $\pm$  8.27e-06 & 9.27743e-02 &      9.27544e-02 \\
   6 &  8.990e-02    &  8.52115e-02  $\pm$  8.73e-06 & 8.52048e-02 &      8.51770e-02 \\
   7 &  8.107e-02    &  7.93875e-02  $\pm$  9.10e-06 & 7.93811e-02 &      7.93436e-02 \\
   8 &  7.400e-02    &  7.47551e-02  $\pm$  9.40e-06 & 7.47496e-02 &      7.47012e-02 \\
   9 &  6.820e-02    &  7.09844e-02  $\pm$  9.64e-06 & 7.09795e-02 &      7.09191e-02 \\
  10 &  6.334e-02    &  6.78504e-02  $\pm$  9.85e-06 & 6.78464e-02 &      6.77734e-02 \\
  11 &  5.921e-02    &  6.52055e-02  $\pm$  1.00e-05 & 6.52021e-02 &      6.51163e-02 \\
  12 &  5.563e-02    &  6.29418e-02  $\pm$  1.02e-05 & 6.29389e-02 &      6.28404e-02 \\
  13 &  5.252e-02    &  6.09835e-02  $\pm$  1.03e-05 & 6.09806e-02 &      6.08695e-02 \\
  14 &  4.976e-02    &  5.92718e-02  $\pm$  1.04e-05 & 5.92692e-02 &      5.91456e-02 \\
  15 &  4.732e-02    &  5.77638e-02  $\pm$  1.06e-05 & 5.77617e-02 &      5.76255e-02 \\
\end{tabular}
\end{center}
\caption{ Bound obtained by integrating numerically over the right
  hand side of Eq.\ref{eq:9}  
  compared with $\bar g(n)$ for a $40\times40$ lattice and different temperatures.}
\label{table1}
\end{table}

There are three type of corrections to the data of table \ref{table1}, which need to be taken into 
account, but which, as we shall show in the following, do not change the above  conclusion of a crossing of
the curves at $n=8$:\\

(i) effects of finite temperature, 

(ii) effects of the finiteness of the system,

(iii) statistical errors.\\

In the following we comment on how these corrections modify the data.
 
(i) The  Quantum Monte Carlo data presented are at $T\geq 0.005$. The overall 
effect of finite temperature is to
lower the absolute value of the correlations and therefore also the 
value of the $\bar g(n)$.  
The effect of finite temperature is to
shift the crossing of the bound and  $\bar g(n)$ 
to larger n, or eventually to destroy a crossing completely.

The functional dependence of the internal energy  $U(T)$, which up to an overall factor
$3 z$ ($z=2$ is the coordination number of the two-dimensional square lattice)
equals the correlation-function at distance one, has been determined for low
$T$ by spin wave theory \cite{Kubo,Oguchi} as
\begin{eqnarray}
\label{eq:10}
U(T)=-e_0 +b T^3.
\end{eqnarray}
The  coefficient is given in \cite{Taka} as 
$b=\frac{\zeta(3)}{2e_0 \pi}\approx 0.2853626$, 
so the correction for distance one 
is  $\approx \frac{b}{6} 10^{-7}$, which is two orders of magnitude smaller than the statistical
error, (see point (iii)). 

For distances larger than one, we fitted the data as a function of temperature
(taking the exponent of T as fit parameter)
for $T=0.005,0.025,0.05,0.075$
and found the corrections due to finite temperature all of the order of $ 10^{-5}$, which is the order of the
statistical error. Therefore we do not give any finite temperature corrections.

(ii) The absolute value of the correlations 
in the thermodynamic limit
are smaller than in systems of finite size. This means that the effect of finite system size is opposite to the effect
of temperature.
The finite size behaviour of the ground state energy 
is well studied for the Heisenberg model on the square lattice.
Arguments originating from the quantum nonlinear sigma model description \cite{CHN} of the Heisenberg model 
to lowest order in system size give
\begin{eqnarray}
\label{eq:11}
-e_0=-e_0(N) +\frac{c} {N^3},\ \ \text{with} \ \ c>0
\end{eqnarray}

where $-e_0(N)$ is the ground state energy of a system of size $N\times N$.
Though the corrections are not  substantial, they do effect the 
results, and taking into account, 
that the finite size errors in contrast to the finite temperature 
effects, might falsely lead to a crossing, we extrapolated the data for
$N=24...40$ using the functional  dependence Eq.\ref{eq:11}, which we found
well satisfied also for larger distances. 
The results are shown in table \ref{table2}.
One sees that the numeric values are changed but the crossing point is 
still at $n=8$.
\begin{table}
\begin{center}
\begin{tabular}{c|c|c|c|c|}
        n& Bound & $T=0.025  $ & T=0.025 extrapolated \\\hline

   1 &  2.297e-01     & 1.80794e-01 &       1.80791e-01  $\pm$ 5.09e-06 \\
   2 &  1.714e-01     & 1.40302e-01 &       1.40295e-01  $\pm$ 7.87e-06 \\
   3 &  1.383e-01     & 1.17678e-01 &       1.17668e-01  $\pm$ 9.53e-06 \\
   4 &  1.166e-01     & 1.02997e-01 &       1.02983e-01  $\pm$ 1.07e-05 \\
   5 &  1.013e-01     & 9.27743e-02 &       9.27534e-02  $\pm$ 1.15e-05 \\
   6 &  8.990e-02     & 8.52048e-02 &       8.51762e-02  $\pm$ 1.21e-05 \\
   7 &  8.107e-02     & 7.93811e-02 &       7.93428e-02  $\pm$ 1.26e-05 \\
   8 &  7.400e-02     & 7.47496e-02 &       7.46995e-02  $\pm$ 1.30e-05 \\
   9 &  6.820e-02     & 7.09795e-02 &       7.09154e-02  $\pm$ 1.34e-05 \\
  10 &  6.334e-02     & 6.78464e-02 &       6.77663e-02  $\pm$ 1.37e-05 \\
  11 &  5.921e-02     & 6.52021e-02 &       6.51035e-02  $\pm$ 1.39e-05 \\
  12 &  5.563e-02     & 6.29389e-02 &       6.28188e-02  $\pm$ 1.41e-05 \\
  13 &  5.252e-02     & 6.09806e-02 &       6.08346e-02  $\pm$ 1.43e-05 \\
  14 &  4.976e-02     & 5.92692e-02 &       5.90923e-02  $\pm$ 1.45e-05 \\
  15 &  4.732e-02     & 5.77617e-02 &       5.75473e-02  $\pm$ 1.46e-05 \\
\end{tabular}
\end{center}
\caption{ Bound obtained by integrating numerically over the right
  hand side of Eq.\ref{eq:9}  
  compared with $\bar g(n)$ extrapolated for N=40,36,32,24  at $T=0.025$.}
\label{table2}
\end{table}

(iii)  We compute $\Delta x =\frac{1}{\sqrt{N_{MC}}}\sqrt{\langle x^2 \rangle - \langle x\rangle^2 }$
 (where the observable x stands for the value of a correlation at a
given distance, temperature and system size and $N_{MC}$ is the number of
 Monte Carlo iterations), 
which is a reliable estimate for the statistical error of the mean value
$\langle x \rangle$, since for the algorithm of Ref.\cite{BW} the autocorrelation time is of 
order one and the Monte Carlo configurations are almost independent.
To assess the quality of our error analysis we also returned to the case of the one-dimensional
antiferromagnetic Heisenberg model ( see Appendix ) and compared results with
independent streams of random numbers.

To calculate an upper limit to the errors of $\bar g(n)$, the errors of the correlations
where added up ( being evaluated with the same configurations, they 
are not independent).

To conclude, the error analysis shows that the short range correlations
entering  Eq.\ref{eq:8} were determined with sufficiently high accuracy 
to prove the existence of a crossing of the bound and the
Quantum Monte Carlo data for $\bar g(n)$ at $n=8$ and therefore 
to show the existence of long range order.\\

{\bf Appendix}

(1) In this Appendix we list the correlations 
of a one-dimensional Heisenberg model with periodic boundary conditions 
%\begin{eqnarray}
%H=  \sum_{<xy>} \vec S_x \vec S_y  \nonumber
%\end{eqnarray}
and chain length $N=400$  at T=0.005 compared with 
results of Ref.\cite{SST} for infinite chain length and $T=0$.
\begin{table}
\begin{center}
\begin{tabular}{c|c|c|c|c|}
& &\\
 Distance & Quantum Monte Carlo & Bethe-Ansatz\\ \hline
   0 &  0.25000000 (  0) \   & \ \\                 
   1 & -0.14771586 (198) \   & -0.1477157268 \ \\  
   2 &  0.06067787 (324) \   &  0.0606797699 \ \\  
   3 & -0.05024194 (282) \   & -0.0502486272 \ \\  
   4 &  0.03464515 (281) \   &  0.0346527769 \ \\  
   5 & -0.03088096 (260) \   & -0.0308903666 \ \\  
   6 &  0.02443619 (255) \   &  0.0244467383 \ \\  
   7 & -0.02248413 (242) \   & -0.0224982227 \ \\  
   8 &  0.01895736 (236) \   &  \\                
\end{tabular}
\end{center}
\caption{Correlations for a chain with $N=400$ sites at T/J=0.005 compared with 
results from Ref. \cite{SST}.}
\label{table0}
\end{table}

For  the internal energy $U(T)$ of the Heisenberg {\sl chain} 
the temperature dependence for low $T$
is $ U(T)=-e_0^1 +a T^2 $
with the ground-state energy $e_0^1=0.4431471804$ for 400 sites and 
$e_0^1=-\frac{1}{4} + \ln 2 $ for the infinite size system\cite{Hulthen}. 
and the coefficient $a=\frac{1}{3}$ given in Ref. \cite{Babujian,Affleck}.
This means that the correction for the correlations
in table\ref{table0} due to finite temperatures are of the order of $10^{-5}$.

(2) The exact values of the correlation functions \cite {DGHK,Dam}
for distance one and two at $T=0$ for a chain with 400 sites are 
$\langle S^3_0 S^3_1\rangle_{400}= -0.147717441765735$
and $\langle S^3_0 S^3_2\rangle_{400}=  0.0606813790491800$.
The  above data show that the error analysis concerning statistical errors
and finite temperature effects is consistent.\\

{\bf Acknowledgement}

I am indebted to  Prof.~E.H.~Lieb for bringing 
the problem of longrange order to my attention and for his interest in this work.

\end{document}